\documentclass[11pt,twoside]{article}
\usepackage{asp2010}

\resetcounters

\begin{document}

\title{On the origin of high-field magnetic white dwarfs}
\author{E. Garc\'\i a--Berro,$^{1,2}$ 
        S. Torres,$^{1,2}$ 
        P. Lor\'en--Aguilar,$^{1,2}$ 
        G. Aznar--Sigu\'an,$^{1,2}$
        J. Camacho,$^{1,2}$
        B. K\"ulebi,$^{2,3}$ 
        J. Isern,$^{2,3}$ 
        L.G. Althaus,$^{4}$ and
        A.H. C\'orsico$^4$} 
\affil{$^1$Departament de F\'\i sica Aplicada, 
           Universitat Polit\`ecnica de Catalunya,  
           c/Esteve Terrades, 5,  
           08860 Castelldefels,  
           Spain}
\affil{$^2$Institute for Space Studies of Catalonia,
           c/Gran Capit\`a 2--4, Edif. Nexus 104,   
           08034  Barcelona, 
	   Spain}
\affil{$^3$Institut de Ci\`encies de l'Espai (CSIC), 
           Facultat de Ci\`encies, 
           Campus UAB, 
           Torre C5-parell, 
           08193 Bellaterra, 
           Spain}
\affil{$^4$Facultad de Ciencias Astron\'omicas y Geof\'{\i}sicas, 
           Universidad Nacional de La Plata, 
           Paseo del Bosque s/n, 
           (1900) La Plata, 
           Argentina}

\begin{abstract}
High-field magnetic  white dwarfs have  been long suspected to  be the
result of stellar mergers. However, the nature of the coalescing stars
and the precise  mechanism that produces the magnetic  field are still
unknown.   Here  we  show  that the  hot,  convective,  differentially
rotating  corona present in  the outer  layers of  the remnant  of the
merger of two  degenerate cores is able to  produce magnetic fields of
the required strength that do  not decay for long timescales.  We also
show,  using  an  state-of-the-art  Monte Carlo  simulator,  that  the
expected number  of high-field magnetic white dwarfs  produced in this
way is consistent with that found in the solar neighborhood.
\end{abstract}

\section{Introduction}

High-field  magnetic white  dwarfs  have field  strengths larger  than
10$^6$~G  and  up  to  10$^9$~G \citep{Schmidtetal03}.   The  galactic
population   of   these   white   dwarfs   exhibits   two   remarkable
peculiarities.  The  first one is  that very few  of them belong  to a
non-interacting  binary system  \citep{Kawkaetal07}, while  the second
one    is    that    they    are    more    massive    than    average
\citep{Silvestrietal07}.  Generally  speaking there are  two competing
scenarios to explain the formation of these stars.  One possibility is
that these  white dwarfs  descend from single  stars, so  the magnetic
field is a fossil of previous evolution \citep{Angeletal81}.  However,
this scenario cannot explain  why they are preferentially massive, and
why they  are not found in non-interacting  binary systems.  Recently,
it has  been suggested \citep{Toutetal08,  Nordhausetal11} that strong
magnetic fields can be produced  during a common envelope episode in a
close  binary system  in which  one of  the components  is degenerate.
During this phase, spiral-in of the secondary in the extended envelope
results in differential rotation  in the convective envelope, which in
turn  produces a  stellar  dynamo  which could  give  raise to  strong
magnetic fields.  However,  it has been shown that  the magnetic field
produced  in this  way  does not  penetrate  in the  white dwarf,  but
instead   decays  rapidly   when  the   common  envelope   is  ejected
\citep{PotteryTout10}.

The coalescence of  double degenerate objects has been  the subject of
numerous   theoretical   studies  during   the   last  years.    Large
computational efforts  have been  devoted to study  this intrinsically
three-dimensional problem because it is  thought to be at the heart of
several  interesting astrophysical phenomena,  among which  we mention
Type   Ia  supernova   \citep{Webbink84,   IbenyTutukov84},  magnetars
\citep{Kingetal01},  and   hydrogen-deficient  carbon  stars   and  of
R~Corona Borealis stars \citep{Longlandetal11, Longlandetal12}.  Also,
the  large  metal abundances  found  around  some hydrogen-rich  white
dwarfs with  dusty disks  around them could  also be explained  by the
merger    of   a    carbon-oxygen   and    a   helium    white   dwarf
\citep{Garciaberroetal07}.  Finally, during  the phase previous to the
coalescence of a double  white-dwarf close binary system gravitational
waves are  emitted, and it  has been shown  that LISA will be  able to
detect them  \citep{Lorenaguilaretal05}.  Here we  quantitatevely show
that the merger of two  degenerate cores can also explain the presence
of  very high  magnetic fields  in some  white dwarfs  ---  a scenario
initially suggested by \cite{WickramasingheyFerrario00}.

\section{An $\alpha\omega$ dynamo in the merger remnant}

Full three-dimensional  simulations of the merger of  two white dwarfs
\citep{Guerreroetal04, Lorenaguilaretal09}  have shown that  the final
remnant of the coalescence is a central white dwarf containing all the
mass of the undisrupted primary. On top of this white dwarf there is a
hot  corona  made  of  about   half  of  the  mass  of  the  disrupted
secondary.  Finally,  surrounding this  structure  a rapidly  rotating
Keplerian disk can be found. This disk contains nearly all the mass of
the secondary which has not been incorporated to the hot corona, since
little  mass ($\sim 10^{-3}\,  M_{\sun}$) is  ejected from  the system
during the  merger process.  The structure of  the system is  shown in
Fig.~1.

\begin{figure}[t]
\vspace{9.5cm}
\begin{center} 
\includegraphics{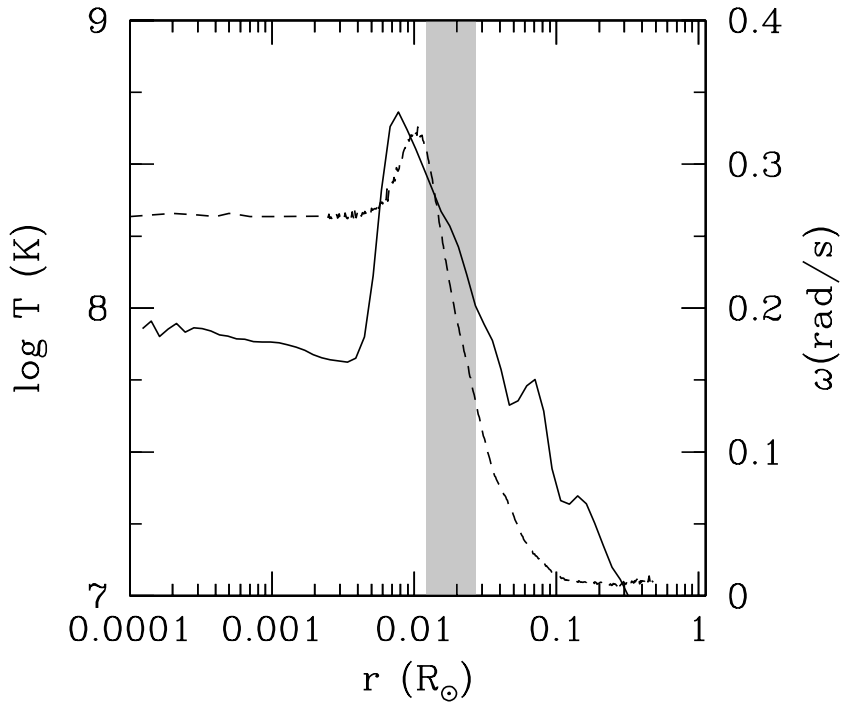} 
\caption{Temperature (solid line,  left scale) and rotational velocity
  (dashed line, right scale) stratifications in the final remnant of a
  double white dwarf merger in  which the binary system is composed of
  two stars  of 0.6 and  $0.8\, M_{\sun}$. The central  spinning white
  dwarf rotates as a rigid body with a rotational velocity $\omega\sim
  0.26$~s$^{-1}$, and  the corona rotates differentially,  with a peak
  rotation  rate $\omega\sim  0.33$~s$^{-1}$.  These  velocities arise
  from energy and angular  momentum conservation, since little mass is
  ejected  from the  system, so  the orbital  angular momentum  of the
  binary  system is  invested in  spinning up  the remnant,  while the
  available energy  is primarily invested in heating  the corona.  The
  location of the convective region is displayed by the shaded area.}
\end{center}
\label{corona} 
\end{figure} 

The temperature  gradient in the  hot corona is  high, and thus  it is
convective. Using the Schwarzschild  criterion we found that the inner
and  outer edges  of the  convective region  are located  at $R\approx
0.012\, R_{\sun}$, and $R \approx 0.026\, R_{\sun}$, respectively, and
that the  total mass  inside this region  is $\sim 0.24  \, M_{\sun}$.
Assuming  energy equipartition,  the  resulting $\alpha\omega$  dynamo
produces  a magnetic field  $B^2/8\pi\approx \rho(\omega  R)^2/2$. For
the  typical values  found in  our simulations  the magnetic  field is
$B\sim 3.2\times  10^{10}$~G. For this  mechanism to be  efficient the
dynamo must work for several convective turnovers before the energy of
the hot corona is radiated away. The temperature of the corona is very
high,  and hence it  is preferentially  cooled by  neutrinos.  Typical
luminosities  are  $L_\nu\sim 4.0\times  10^2\,  L_{\sun}$, while  the
total thermal  energy of the corona is  $U\sim 8.8\times 10^{48}$~erg.
Consequently,  the  convective  shell  lasts for  $\tau_{\rm  hot}\sim
1.8\times 10^5$~yr.   The convective turnover  timescale is $\tau_{\rm
conv}\approx H_P/v_{\rm  conv}$, where $H_P\approx  2.7\times 10^8$~cm
is the pressure scale  height and $v_{\rm conv}\approx 8.0\times 10^7$
cm/s is  the convective velocity.  Thus,  $\tau_{\rm conv}\sim 3.3$~s,
and during  the lifetime  of the hot  corona the number  of convective
cycles is  sufficiently large.  Thus, the  $\alpha\omega$ mechanism is
able to produce strongs magnetic fields.

Now that we have proved  that large magnetic fields can be established
in the  aftermath of  a double degenerate  merger, it is  necessary to
assess  if  during the  lifetime  of  the  resulting white  dwarf  the
magnetic field diffuses outwards, to the surrounding disk, or inwards,
to the degenerate primary. Thus,  we solved the diffusion equation for
both the disk and the degenerate core, and we found that the timescale
for  diffusion of  the magnetic  field across  the disk  is $\tau_{\rm
disk}\sim 2.0\times 10^{11}$~yr, while for the diffusion timescale for
the central white  dwarf turns out to be  $\tau_{\rm WD}\sim 4.3\times
10^9$~yr.  Thus, it  can safely stated that as  the white dwarf cools,
the magnetic field is not allowed to diffuse across the keplerian disk
nor  to penetrate  in the  degenerate core,  and  consequently remains
confined to the surface layers.

Our model predicts that the masses of high-field magnetic white dwarfs
should be larger than the average  of field white dwarfs and that they
should be  observed as single white dwarfs,  as observationally found.
However, high-field  magnetic white dwarfs  are generally found  to be
slow  rotators \citep{WickramasingheyFerrario00}.   Nevertheless, this
apparent  shortcoming  can be  easily  solved.   If  the rotation  and
magnetic axes  are misaligned, magneto-dipole  radiation rapidly spins
down  the  white dwarf.   The  evolution  of  the rotational  velocity
\citep{Benacquistaetal03} is given by $\dot{\omega} = -2\mu^2 \omega^3
\sin^2\alpha /(3Ic^3)  $, where  $I$ is the  moment of inertia  of the
white dwarf, $\alpha$  is the angle between the  magnetic and rotation
axes and $\mu=BR_{\rm WD}^3$.   Adopting typical values resulting from
our  SPH  simulations  we  obtain  a  spin-down  timescale  $\tau_{\rm
  MDR}\sim  2.4\times10^{8}/\sin^2\alpha$~yr,  when  a field  strength
$B=10^{7}$~G is  adopted.  Hence, if  both axes are  perfectly aligned
the remnant of the coalescence  will be a high-field, rapidly rotating
magnetic  white dwarf.   On  the  contrary, if  both  axes are  nearly
perpendicular magneto-dipole radiation efficiently brakes the remnant.
Consequently,   very  young,   hot,   ultramassive,  slowly   rotating
high-field magnetic  white dwarfs can  also be easily  accommodated in
our model.

On the other hand, spectro-polarimetric observations show that in most
cases high-field magnetic white  dwarfs have fields with both toroidal
and  poloidal components.   Our  scenario can  also  account for  this
observational fact, since  in the $\alpha\omega$ mechanism, convection
is responsible for the generation of poloidal fields, whereas rotation
is responsible  for the generation of toroidal  fields. In particular,
the energy available to generate the poloidal field component is $\rho
v_{\rm conv}^2/2\sim  4.0\times 10^{20}$~erg, which is  $\sim 10\%$ of
the  energy available  to build  the toroidal  component, $\rho(\omega
R)^2/2\sim 5.5\times 10^{21}$~erg.  Thus,  we expect that the magnetic
field geometry of the remnant of the merger will be complex.

\section{The number of high-field magnetic white dwarfs in the solar
neighborhood}

To  assess  the  number of  mergers  that  could  occur in  the  solar
neighborhood    we   expanded   an    existing   Monte    Carlo   code
\citep{Garciaberroetal99, Torresetal02, Garciaberroetal04} designed to
study the galactic population of single white dwarfs to deal with that
of double  degenerates. The population synthesis code  is described in
full  depth  in  \cite{Garciaberroetal12},  where  all  the  necessary
ingredients are detailed.  Here, for  the sake of conciseness, we only
present the most significant results of our calculations.

Our  population synthesis  calculations predict  that the  fraction of
merged  double degenerate  cores in  the solar  neighborhood  is $\sim
2.9\%$ of the  total population.  This number includes  not only white
dwarf  mergers ($\sim  0.3\%$), but  also the  coalescence of  a white
dwarf and a giant star with  a degenerate core ($\sim 1.1\%$), and the
merger of two  giants with degenerate cores ($\sim  1.5\%$).  In these
two last cases the mergers  obviously occur during the common envelope
phase,  while  for  the  cases  in which  two  white  dwarfs  coalesce
gravitational  wave  radiation is  the  final  driver  of the  merger.
Finally, it is  important to mention as well  that the distribution of
remnant  masses  is  nearly  flat,  in  agreement  with  the  observed
distribution     of     masses     of    magnetic     white     dwarfs
\citep{NaleztyyMadej04}.

Within   20~pc    of   the   Sun   there   are    122   white   dwarfs
\citep{Holbergetal08},     and    contains    14     magnetic    stars
\citep{Kawkaetal07}.  Although  scarce, this sample  is 80\% complete,
and allows reliably determine the true incidence of magnetism in white
dwarfs.  Mass determinations are available for 121 of them.  Of the 14
magnetic white dwarfs  in local sample, 8 have  magnetic fields larger
than $10^7$~G,  and 3 have masses  larger than $0.8\,  M_{\sun}$ --- a
value which  is $\sim 2.5\sigma$ away  from the average  mass of field
white dwarfs. This  mass cut is chosen to cull  only white dwarfs that
are  expected to  be the  result  of stellar  mergers.  Actually,  our
calculations  predict that  $\sim 4$  white dwarfs  are the  result of
double  degenerate  mergers,  and   have  masses  larger  than  $0.8\,
M_{\sun}$,   in  agreement   with  observations.    Additionally,  our
simulations  predict that the  fraction of  white dwarfs  more massive
than $\sim 0.8\, M_{\sun}$  resulting from single stellar evolution is
$\sim  10\%$.   Consequently, the  expected  number  of massive  white
dwarfs in  the local sample should  be $\sim 12$.   Instead, the local
sample contains 20, pointing  towards a considerable excess of massive
white dwarfs, which could be the  progeny of mergers.  The rest of the
population of  magnetic white dwarfs  ($\sim 5$) --- those  with small
magnetic field strengths --- would be well the result of the evolution
of single stars \citep{AznarCuadradoetal04}.

\section{Conclusions}

It has  been shown that  the hot, convective,  differentially rotating
corona   predicted  by   detailed   Smoothed  Particle   Hydrodynamics
simulations  of the coalescence  of two  degenerate stellar  cores can
produce  very high  magnetic fields.   We have  also shown  that these
magnetic fields are confined to the outer layers of the remnant of the
coalescence, and do not propagate neither to the interior of the white
dwarf or  to the debris region.   Our model has  two clear advantages,
which meet the requirements imposed  by observations. The first one is
that  it  naturally predicts  that  high-field  magnetic white  dwarfs
should preferentially  have high masses. The second  advantage is that
it  explains why  high-field magnetic  white  dwarfs are  found to  be
single stars.   Moreover, our scenario  predicts that if  the rotation
and magnetic  axes are  not aligned the  magnetic white  dwarf rapidly
spins down due  to the emission of magnetodipole  radiation.  Thus, we
expect  that high-field  magnetic white  dwarfs should  have different
rotation periods,  depending on their  evolutionary status and  on the
geometry of  the magnetic field.  Moreover,  in the case  in which the
masses of the merging objects white dwarfs are not equal the keplerian
disk  can  eventually   form  a  second-generation  planetary  system.
Disruption of small bodies  in this planetary system could contaminate
the atmospheric layers  of the magnetic white dwarf.   This finding is
important,  as it can  explain the  recently discovered  population of
metallic magnetic  white dwarfs.  If,  on the contrary, the  masses of
the  merging  white  dwarfs  are  similar the  remnant  has  spherical
symmetry  and rotates  very rapidly,  as observed  in  some high-field
magnetic  white dwarfs.  Also, the  geometry of  the  surface magnetic
fields  can be well  explained by  our model.   Finally, we  have also
shown that the expected number of double degenerate mergers is roughly
consistent with the number of  high-field magnetic white dwarfs in the
local sample.   In summary, our  calculations indicate that  a sizable
fraction of all  high-field magnetic white dwarfs could  be the result
of double  degenerate mergers, a hypothesis  previously anticipated by
\cite{WickramasingheyFerrario00},  but  not  hitherto demonstrated  by
a quantitative analysis.

\acknowledgements 
This research  was supported by AGAUR, by  MCINN grants AYA2011--23102
and  AYA08-1839/ESP, by  the European  Union FEDER  funds, by  the ESF
EUROGENESIS  project (grant  EUI2009-04167), by  AGENCIA:  Programa de
Modernizaci\'on  Tecnol\'ogica BID 1728/OC-AR,  and by  PIP 2008-00940
from CONICET.

\bibliography{12eg}

\end{document}